\documentclass[%
 aip,
 amsmath,amssymb,
 reprint,%
]{revtex4-1}

\usepackage{graphicx}
\usepackage{dcolumn}
\usepackage{bm}
\usepackage{hyperref}
\hypersetup{
    colorlinks=true,
    linkcolor=blue,
    filecolor=blue,      
    urlcolor=blue,
    pdfpagemode=FullScreen,
    }
\usepackage[utf8]{inputenc}
\usepackage[T1]{fontenc}
\usepackage{mathptmx}
\usepackage{etoolbox}
\usepackage{graphicx}
\usepackage{units}
\usepackage{placeins}
\usepackage{float}
 \usepackage{url}

\makeatletter
\def\@email#1#2{%
 \endgroup
 \patchcmd{\titleblock@produce}
  {\frontmatter@RRAPformat}
  {\frontmatter@RRAPformat{\produce@RRAP{*#1\href{mailto:#2}{#2}}}\frontmatter@RRAPformat}
  {}{}
}%
\makeatother
\begin{document}

\preprint{AIP/123-QED}

    \title{Benchmarking energy consumption and latency for neuromorphic computing in condensed matter and particle physics}
\author{Dominique J. K\"osters}\affiliation{University of Twente, Faculty of Science and Technology, P.O. Box 217
7500 AE, Enschede, The Netherlands}\affiliation{Radboud University, Institute for Molecules and Materials, Heyendaalseweg 135, 6525 AJ, Nijmegen, The Netherlands}\affiliation{IBM Research Europe - Z\"urich, 8803 R\"uschlikon, S\"aumerstrasse 4, Switzerland}

\author{Bryan A. Kortman}\affiliation{University of Twente, Faculty of Science and Technology, P.O. Box 217
7500 AE, Enschede, The Netherlands}\affiliation{Nikhef, P.O.Box 41882 1098 XG Amsterdam, The Netherlands}

\author{Irem Boybat}\affiliation{IBM Research Europe - Z\"urich, 8803 R\"uschlikon, S\"aumerstrasse 4, Switzerland}

\author{Elena Ferro}\affiliation{IBM Research Europe - Z\"urich, 8803 R\"uschlikon, S\"aumerstrasse 4, Switzerland}\affiliation{Eidgen\"ossische Technische Hochschule Z\"urich, Department of Information Technology and Electrical Engineering, Gloriastrasse 35, 8092 Z\"urich, Switzerland}

\author{Sagar Dolas}\affiliation{SURF Cooperation, Innovation Team, Moreelespark 48, 3511 EP, Utrecht, The Netherlands }

\author{Roberto Ruiz de Austri}\affiliation{University of Valencia-CSIC, Instituto de F\'isica Corpuscular, Parc Cient\'ific UV, c/ Catedr\'atico Jos\'e Beltr\'an 2, E-46980 Paterna, Spain}

\author{Johan Kwisthout}\affiliation{Radboud University, Donders Institute for Brain, Cognition and Behaviour, P.O. Box 9104 6500 HE, Nijmegen, The Netherlands}

\author{Hans Hilgenkamp}\affiliation{University of Twente, Faculty of Science and Technology, P.O. Box 217
7500 AE, Enschede, The Netherlands}\affiliation{University of Twente, MESA+ Institute for Nanotechnology,  P.O. Box 217
7500 AE Enschede, The Netherlands}

\author{Theo Rasing}\affiliation{Radboud University, Institute for Molecules and Materials, 6525 AJ Nijmegen, Heyendaalseweg 135, The Netherlands}

\author{Heike Riel}\affiliation{IBM Research Europe - Z\"urich, 8803 R\"uschlikon, S\"aumerstrasse 4, Switzerland}

\author{Abu Sebastian}\affiliation{IBM Research Europe - Z\"urich, 8803 R\"uschlikon, S\"aumerstrasse 4, Switzerland}

\author{Sascha Caron}\affiliation{Nikhef, P.O.Box 41882 1098 XG Amsterdam, The Netherlands}\affiliation{Radboud University, Institute for Mathematics, Astrophysics and Particle Physics, Heyendaalseweg 135, 6525 AJ, Nijmegen, The Netherlands}

\author{Johan H. Mentink}
 \altaffiliation[ ]{Corresponding Author: Johan H. Mentink, J.Mentink@science.ru.nl}
\affiliation{Radboud University, Institute for Molecules and Materials, 6525 AJ Nijmegen, Heyendaalseweg 135, The Netherlands}


\date{\today}

\begin{abstract}
The massive use of artificial neural networks (ANNs), increasingly popular in many areas of scientific computing, rapidly increases the energy consumption of modern high-performance computing systems. An appealing and possibly more sustainable alternative is provided by novel neuromorphic paradigms, which directly implement ANNs in hardware. However, little is known about the actual benefits of running ANNs on neuromorphic hardware for use cases in scientific computing. Here we present a methodology for measuring the energy cost and compute time for inference tasks with ANNs on conventional hardware. In addition, we have designed an architecture for these tasks and estimate the same metrics based on a state-of-the-art analog in-memory computing (AIMC) platform, one of the key paradigms in neuromorphic computing. 
Both methodologies are compared for a use case in quantum many-body physics in two dimensional condensed matter systems and for anomaly detection at 40 MHz rates at the Large Hadron Collider in particle physics. We find that AIMC can achieve up to one order of magnitude shorter computation times than conventional hardware, at an energy cost that is up to three orders of magnitude smaller. This suggests great potential for faster and more sustainable scientific computing with neuromorphic hardware.
\end{abstract}

\maketitle

\section{\label{sec:level1}Introduction}
The energy demand of modern high-performance computing systems is currently rapidly rising owing to the massive use of artificial neural networks (ANNs)\cite{mehonic2022nature}. Moreover, for several of the most challenging compute tasks in scientific computing, applications of ANNs offer competitive advantages over standard algorithms. 
This includes various examples in condensed matter physics\cite{butler2018machine,carleo2019machine,bedolla2020machine} and in particle physics\cite{radovic2018machine,feickert2021living,karagiorgi2022machine}, which features data rates as high as 1 Pb/second in bursts separated by just 25 nanoseconds\cite{Govorkova}. Such workloads are very challenging for conventional hardware since the corresponding energy consumption is simply too high, even for inference tasks with pre-trained networks.

Neuromorphic hardware offers great potential as accelerator for such highly demanding compute tasks. For example, spiking neural networks are considered advantageous for optimization problems\cite{davies2021ieee, schuman2022opportunities, aimone2017neural, fonseca2017using} and have recently also been applied to entangled quantum states\cite{czischek2022scipost}. Another efficient realization of ANNs involves physically instantiating the synaptic weights in memory devices and exploiting the physical attributes of these memory devices to implement the ANN in hardware \cite{Y2020sebastianNatureNano,Y2020yaoNature,Y2022fickISSCC}. This approach, typically referred to as analog in-memory computing (AIMC), would obviate the need to shuttle millions of synaptic weights between the memory and processing units and could lead to significant gains in energy efficiency and latency. However, rather little is known about the actual benefits of running ANNs on AIMC for concrete physics use cases. Moreover, physics users often do not even consider the energy cost of complete ANN workloads as a relevant figure of merit.

In this Letter, we aim to assess the potential of AIMC accelerators for ANN-based use cases in condensed matter and particle physics by benchmarking them against conventional hardware implementations of the ANNs. To this end we develop a generic methodology for measuring the energy cost and compute time for inference tasks with ANNs on CPU and GPU hardware. In addition, we have designed an architecture to estimate these metrics for inference tasks on a Mixed-Precision Analog In-Memory Computing (MP-AIMC) platform. The scientific use cases chosen feature computation of quantum many body states in two dimensions \cite{carleo2017solving,fabiani2019scipost} as well as the 40 MHz challenge for anomaly detection at the Large Hadron Collider (LHC) in particle physics\cite{SCaronDarkMachines,SCaronAnomaly}. By comparing the measurements on conventional hardware against the estimations for MP-AIMC, we find that the latter can reach up to an order of magnitude shorter compute time and down to three orders of magnitude lower energy cost.

\section{Scientific Use cases} 
\subsection{Condensed Matter Physics}
Understanding the effect of correlations on the properties of quantum many-body systems is one of the most fascinating research fields in condensed matter physics. Recently, a new method for the simulation of quantum many body systems has been pioneered inspired by machine learning \cite{carleo2017solving}. In this approach, the many-body wave function is approximated by an ANN and the quantum states generated this way are termed neural-network quantum states (NQS). Already the simplest network, the Restricted Boltzmann Machine (RBM), was found to give competitive advantages over conventional methods, in particular in two dimensions (2D) for which quantum correlations are strongest\cite{carleo2017solving,fabiani2019scipost,nomura2021prx,fabiani2021supermagnonic}.
Despite this potential, the method suffers from long training times when applied to large systems. Moreover, extracting observables from an already trained network takes up a large part of the computational effort. 

\begin{figure}[H]
\centering
\includegraphics[width=0.4\textwidth]{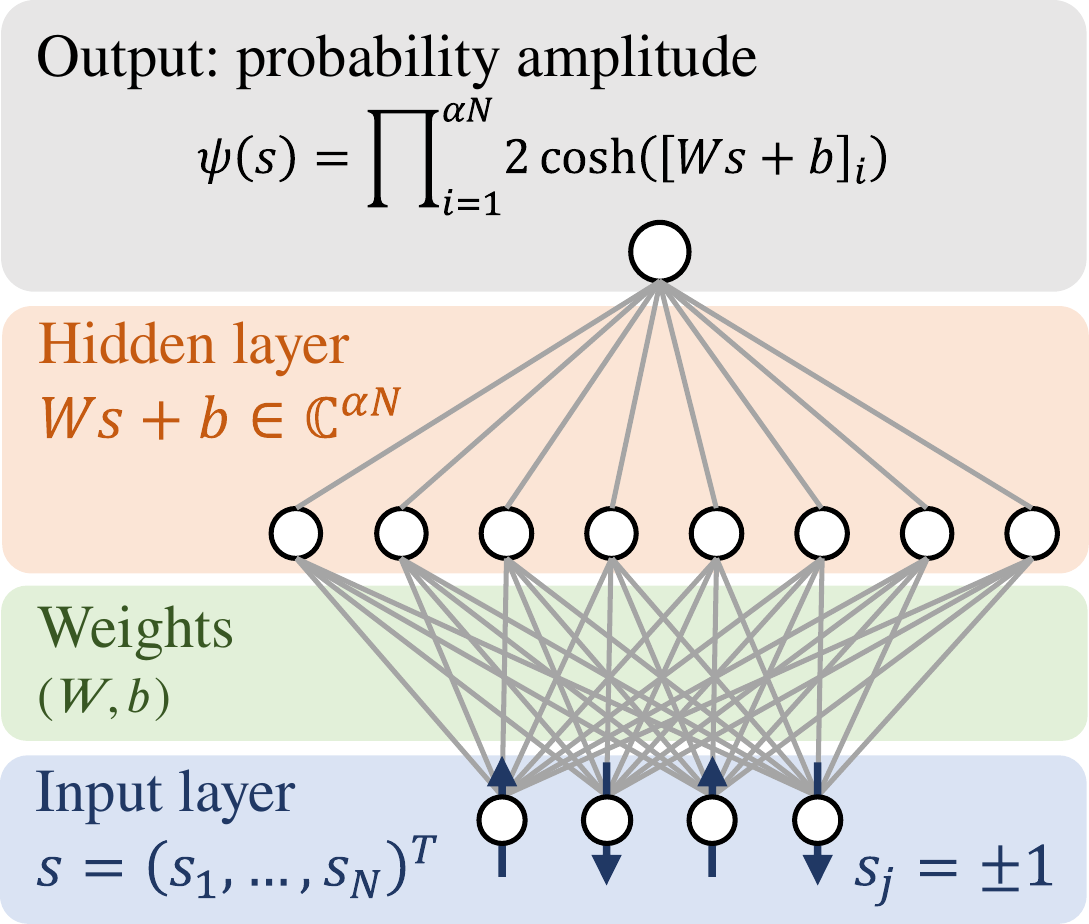}
\caption{Schematic representation of the RBM. the hidden layer size is $\alpha N$, with $\alpha$ a positive integer (in this schematic representation $N=4$ and $\alpha=2$). Each hidden neuron features a hyperbolic cosine activation. Their outputs are multiplied to produce the network output.}
\label{fig: rbm}
\end{figure}

For this use case we consider NQS for the 2D antiferromagnetic Heisenberg model on a square lattice. The Hamiltonian is defined as
\begin{equation}
    \hat{H}=J\sum_{\langle i,j \rangle}\hat{S}_i\cdot \hat{S}_j,
\end{equation}
where $J$ is the exchange constant, $\hat{S}_i$ the quantum spin operators for $S=1/2$ and the sum runs over the nearest neighbours. In the context of NQS, an inference task is defined by evaluation of the RBM wave function, which for translation invariant systems can be written as
\begin{equation}\label{ground_state}
    \psi(s) = \prod_{i=1}^{\alpha N} 2\cosh{\left(\left[Ws + b\right]_i\right)},
\end{equation}
where $\alpha$ is the ratio between the number of spins and hidden layer nodes, $N$ the number of spins in the lattice, $W$ a matrix of dimensions $\alpha N \times N$ with the weights, $b$ a vector of dimension $\alpha N$ with the biases and $s$ represents the input spin configuration, which is encoded in a binary tuple (column vector) $s=(s_1,\ldots,s_N)^T$, with $s_j=\pm1$. 
A schematic representation of the RBM is shown in Figure \ref{fig: rbm}. 
For large systems, $\psi(s)$ is evaluated only for a subset of the $2^N$ possible states by Monte Carlo sampling. For our assessment, it is sufficient to consider a $4\times4$ system for which the input data set is defined by all possible states with zero magnetization. 

In order to obtain ground state properties, the network is trained by minimizing the energy $E = {\langle\psi|\hat{H}|\psi\rangle}/{\langle\psi|\psi\rangle}$ using the stochastic reconfiguration method\cite{Sorella1998prl}, while a closely related training procedure is obtained for quantum dynamics based on the time-dependent variational principle\cite{imada2015prb,carleo2017solving}. In either case, the training itself relies on repeated evaluation of $\psi(s)$ at fixed RBM parameters, before updates are computed. Moreover, evaluation of observables exclusively relies on inference. Therefore, we focus on the inference procedure alone and take already trained weights for the RBM ground state\cite{fabiani2019scipost}. In this case, both the weights and biases can be chosen to be real valued. 

\subsection{Particle Physics}
One of the most important challenges in High Energy Physics (HEP) today is to find rare new physics signals among an abundance of Standard Model (SM) proton-proton collisions. This use case\cite{SCaronAnomaly} uses Deep Learning (DL) techniques in order to find anomalous signals at the LHC among many SM events, also known as anomaly detection. The event rate from which events must be selected at the ATLAS detector is 40 MHz, brought down to the final collection rate of 300Hz using a three staged trigger system. The network in this use case is designed to run on the Level-1 trigger system responsible for reducing the rate to less than 75kHz. Therefore, very low latency is required to ensure the network can keep up with the proton-proton collision rate.

\begin{figure}[H]
\centering
\includegraphics[width=0.4\textwidth]{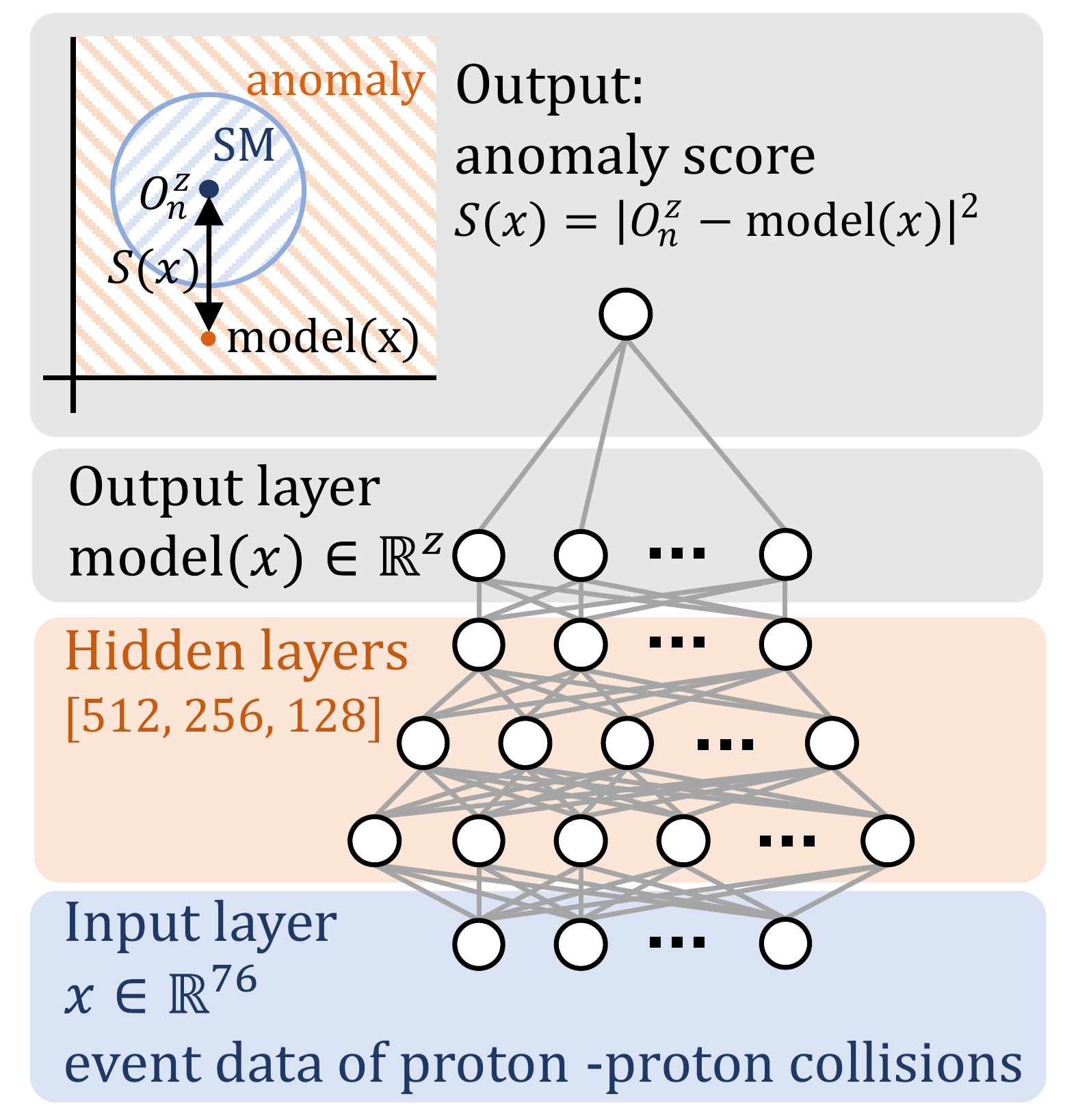}
\caption{A schematic representation of the Deep SVDD model. The input is the event data of proton-proton collisions and the output is a z-dimensional vector denoted model$(x)$. An anomaly score can be given to the event by comparing the model output with a target ($O^z_n$). In this case the model output lays far from the target and the event can therefore be considered an anomaly.}
\label{fig: svdd}
\end{figure}

The network in this use case is designed to detect events that are forbidden in the SM. This is achieved by a One-Class Deep Support Vector Data Description (Deep SVDD) approach, trained to map every input of the Deep SVDD onto a predefined multidimensional point. The distance to this point is then regarded as the final anomaly score.  A schematic representation of the SVDD is shown in Figure \ref{fig: svdd}. The models are trained under the assumption that all SM data falls inside the predetermined manifold. During the testing, the beyond the Standard Model (BSM) data will fall outside this manifold. Several Deep SVDD networks are combined into an ensemble to maximise the efficiency. The networks are fully connected networks which output a constant vector for every input. The loss is defined as
\begin{equation}
    \label{eqn:distancecalculation}
    S(x) = \left[ O^{z}_{n} - \mathrm{model}(x) \right]^2.
\end{equation}

The model maps the input x to the same tensor shape as the vector $O$, with $z$ and $n$ referring to the number of elements and the scalar value, respectively. The activation function of the hidden layers is an exponential linear unit (ELU). In this use case, we employ an ensemble of in total 63 networks. They contain the same network structure but different combination of values [5,8,13,21,34,55,89,144,233] for $z$ and [0,1,2,3,4,10,25] of $n$. 

The data used to test and train this use case consists of 5 data records\cite{Govorkova}. One record containing a mixture of SM processes and four separate records of BSM processes. The data is generated using Pythia 8.240\cite{PYTHIA} using a collision energy of $\sqrt{s}=13$ TeV. The BSM processes were $A\rightarrow{}4l$, $LQ\rightarrow{}b\tau$, $H\rightarrow{}\tau\tau$ and the $H^\pm \rightarrow{}\tau\nu$ process. The detector response is modelled with DELPHES 3.3.2.\cite{DELPHES} Input variables available for each event are the $p_T$, $\eta$ and $\phi$ values of the missing energy, 4 electrons, 4 muons and 10 jets.

\section{Methods}
\subsection{\label{sec:meaurement} Energy-measurement methodology CPU and GPU }
The energy measurements were performed  on the ESC cluster within Innovation Labs at SURF.
The inference of the networks is performed on a dual socket system with 2 Intel Xeon Gold CPU's and a NVIDIA V100 GPU.
The measurements are performed using the Energy Aware Runtime (EAR) software package\cite{EAR} and provides an energy management framework for experimental computing. 
The inference jobs are monitored by measuring the DC node power using a Baseboard Management Controller (BMC), which is a specialized service processor that monitors the physical state of the hardware device using sensors and communicating with the system administrator through an independent connection.
The BMC is part of the Intelligent Platform Management Interface (IPMI), which is a standardized message-based hardware management interface. 

EAR is currently implemented to give the metrics of an entire submitted job on a specific node. The energy metric used employs the DC node energy $E^\text{node}_{DC}$. The metrics presented below are the average energy per inferred sample defined as
\begin{equation}
    E_\text{sample} = \frac{E^\text{node}_{DC}}{N},
\end{equation}
where $N$ is the number of inferred samples, and the throughput $T$ defined as
\begin{equation}
    T = \frac{N}{\Delta t},
\end{equation}
where  $\Delta t$ is the elapsed time. 
An effective latency could be derived from the throughput by taking the inverse, i.e. $L = T^{-1}$.

The samples where supplied in batches to the ANNs. The batch size was chosen to minimize $E_\text{sample}$ by doing a sweep search. The most energy efficient batch size was in these cases also the batch size with the highest throughput. To ensure that initialization processes are insignificant in terms of energy and time, the total computation time of inference is forced to be larger than $99\%$ by increasing the number of inference steps. 
The scripts developed for energy measurements with EAR, applicable for any neural network, are available open access at \href{https://github.com/dkosters/EME}{github}\cite{Kosters2022}.


\subsection{\label{sec:estimation}Energy estimation for mixed-precision analog in-memory computing hardware}
 
In this section we present a dedicated MP-AIMC hardware design, which can support both physics use cases described earlier. Although originating from vastly different fields of physics, the operations necessary to compute the wave function $\psi(S)$ and the anomaly score are similar and therefore the two use cases are easily combined in one design. 

\begin{figure}
    \centering
    \includegraphics[width=0.4\textwidth]{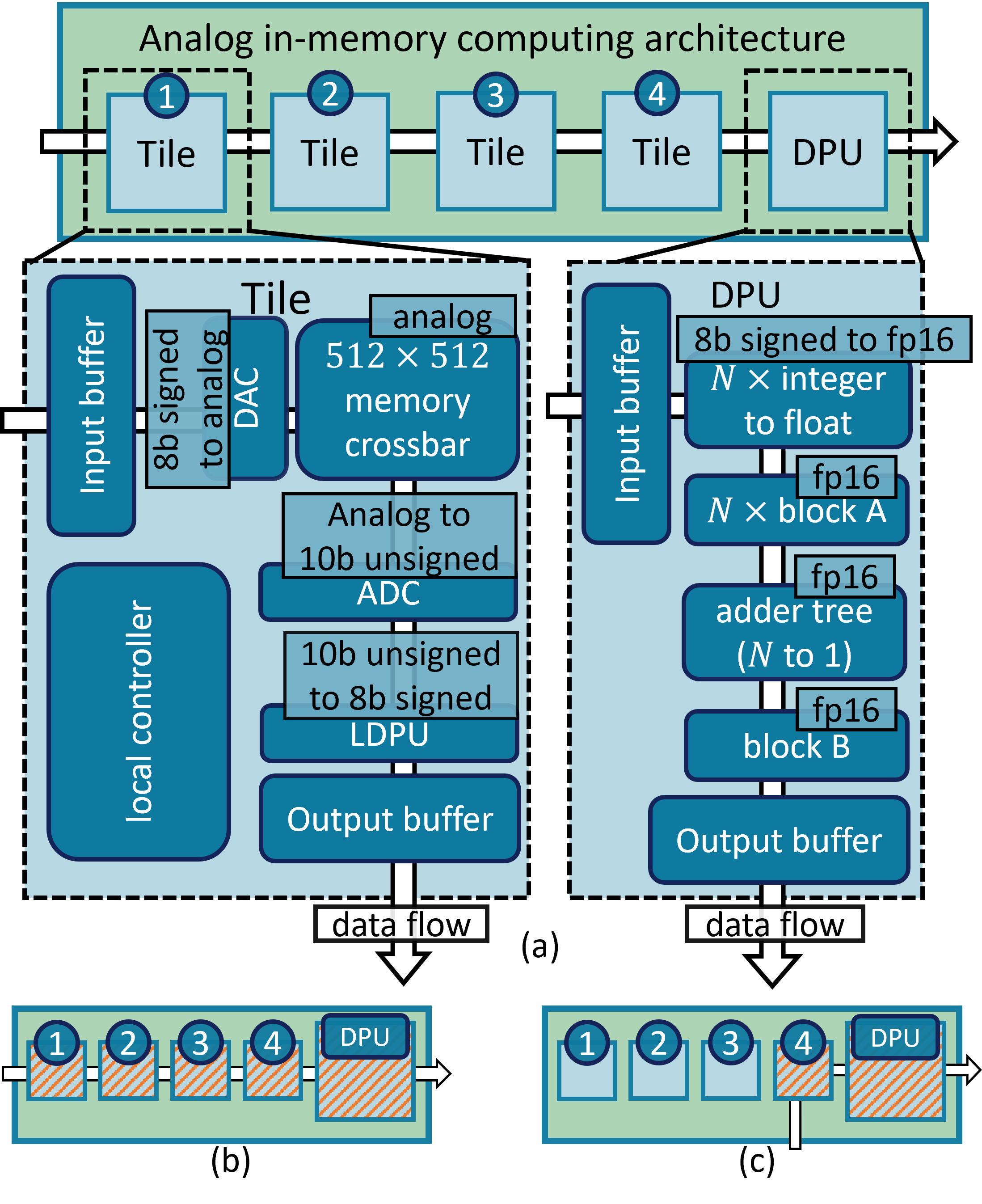}
    \caption{Proposed MP-AIMC architecture. (a) Block diagram of the proposed AIMC-based architecture showing four identical tiles and the DPU. The sub-figure include a zoom-in of one tile and a zoom-in of the DPU where the corresponding data types are mentioned per block. Included in blocks A and B are a hard-wired look-up table, floating point multiplier and floating point adder. Sub-figure (b) shows the utilization of the proposed AIMC-based architecture for the condensed matter physics use case, where only tile 4 and the DPU are utilized. Sub-figure (c) shows the utilization for the particle physics use case, where all 4 tiles and the DPU are utilized.}
    \label{fig:imc_design}
\end{figure}

AIMC hardware is based on memory crossbar arrays with stationary weights, implementing matrix-vector multiplication (MVM) directly in hardware. The proposed MP-AIMC architecture, shown in Figure \ref{fig:imc_design}, is composed of four analog tiles and a Digital Processing Unit (DPU). 
The analog tile is assumed to be similar to a phase-change memory-based AIMC tile designed and fabricated in \unit[14]{nm} CMOS technology node\cite{9696185, narayanan2021fully}. 
The DPU components are also estimated based on circuit designs at the same technology node.

The four tiles are identical and consist of a local controller, 512 Digital-to-Analog Converters (DACs), $512 \times 512$ memory crossbar with Phase Change Memory (PCM) devices, 512 Analog-to-Digital Converters (ADCs) and a Local Digital Processing Unit (LDPU). 

The DAC converts an 8-bit signed integer to a voltage pulse with fixed height (negative and positive numbers have a different fixed height) where the width of the pulse corresponds to the value of the integer. The DACs are connected to the wordlines of the memory crossbar. The crossbar evaluates the vector matrix multiplication by accumulating current along the bitlines according to Ohm’s and Kirchoff’s law. The ADCs are variants of the current-controlled oscillator-based ADCs as described in Ref.       \cite{9696185}. The instantaneous current flowing through the bitlines is digitized and accumulated in a digital counter (which is also part of the ADC) during the application of the input pulse. The ADC outputs two 10-bit unsigned integers, one integer corresponding to positive values and one corresponding to negative values. Due to the compact design of the ADCs, it is possible to operate all 512 ADCs in parallel. The combined latency of the DACs, ADCs and the memory crossbar is estimated to be 40 ns.

The LDPU implements the affine scaling, addition of biases and optionally the ReLU activation function. Due to practical considerations, the Deep SVDD implementation on the MP-AIMC architecture employs a ReLU instead of an ELU activation function.
Both activation functions result in a similar accuracy. 
The LDPU and all the components used inside the DPU are synthesized to verify that both timing and area constraints are met and custom designed for relatively low-precision operations.

The additional DPU takes the network output and computes $\psi(s)$ for NQS and for the SVDD it evaluates the euclidean distance between the target, $O_{n}^{z}$, and the network output, which can be viewed as the anomaly score. This evaluation can be done with the use of look-up tables for activation functions and an adder tree for both use cases. For NQS, only one analog tile and the DPU is utilized (shown in Figure \ref{fig:imc_design} c), while for the SVDD application all 4 tiles and DPU are utilized (Figure \ref{fig:imc_design} b). 

Due to the stationary nature of the weights in the proposed MP-AIMC architecture, this design is well suited for a data-flow in pipeline fashion. This means that the data-flow can be divided in several sequential pipeline stages. 
The throughput is then limited by the slowest pipeline stage, which in this design is estimated to be \unit[50]{ns}, and is independent of network size as long as the individual layers fit on the $512\times512$ crossbar arrays.
When running the workloads of the two use cases with maximum load (entire 512 x 512 crossbar is utilized) the power consumption of the crossbar (the memory array as well as the peripheral circuitry including the data converters) is estimated to be 0.13W, where the peripheral circuitry contributes contributes about 90\% of the power consumption.
The average power consumption of the LDPU and DPU after synthesis are estimated at \unit[0.33]{W} and \unit[0.18]{W} respectively. In this case the average power consumption for the computational and particle physics use cases are \unit[0.63]{W} and \unit[2.03]{W} respectively. Note that, in the former case just one tile and the DPU are used whereas in the later case, all four tiles and the DPU are employed. Energy consumption of communication between tiles and between the final tile and the DPU is included in the energy estimation.

\begin{figure*}
    \centering
    \includegraphics[width=0.9\textwidth]{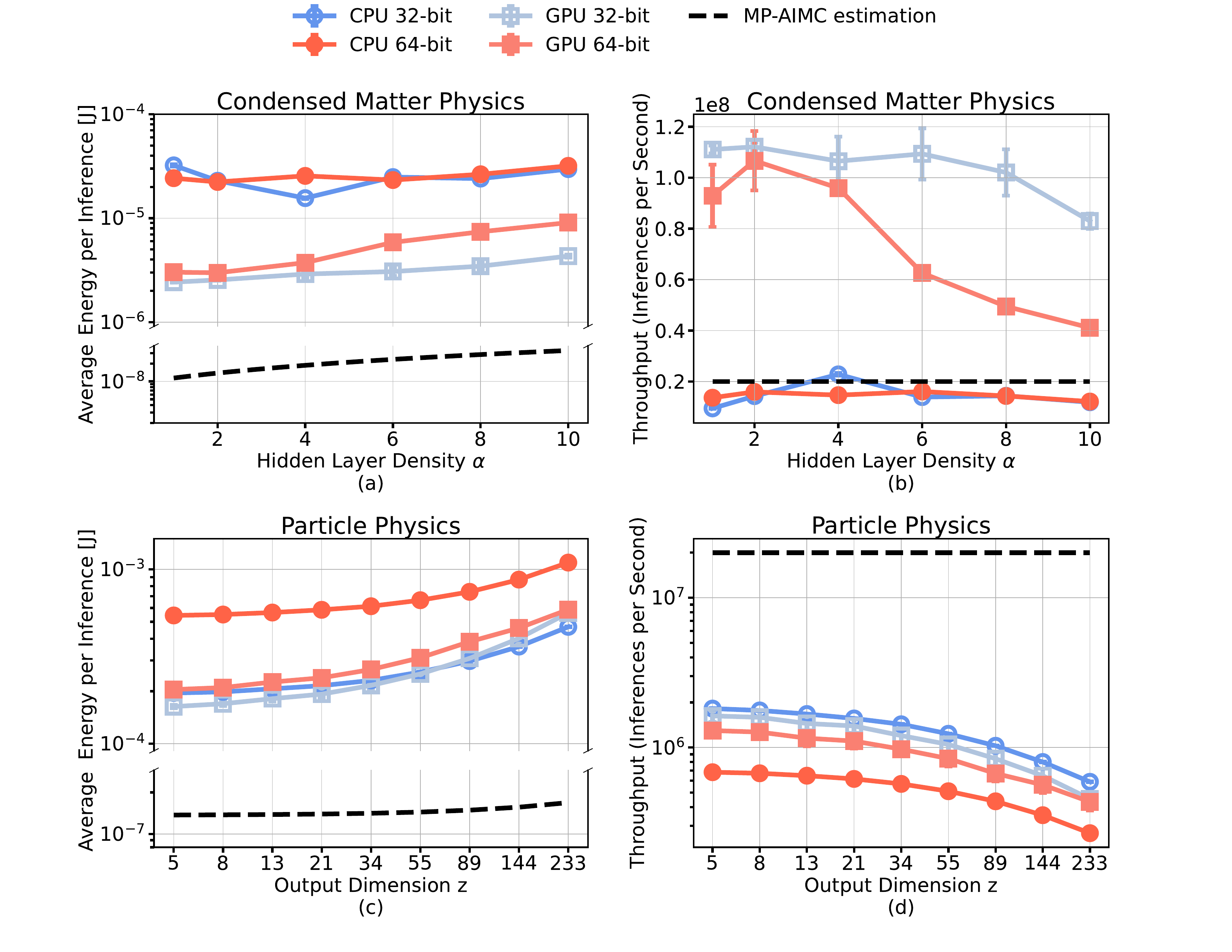}
    \caption{Results benchmarks on CPU (Intel Xeon Gold dual socket) and GPU (NVIDIA V100) and estimations for the proposed AIMC architecture. (a) The average energy per inference (or average energy per state) in Joule and (b) the throughput inferences per second (or states per second) for the condensed matter physics use case. (c) The average energy per inference (or average energy per event) in Joule and (d) the throughput inferences per second (or events per second) for the particle physics use case.
    The measurements on the CPU and GPU are repeated 10 times and the standard deviation is included in the plots. 
    The batch size used for the condensed matter physics use case is equal to all states with zero magnetization ($12870$) and the batch size used for the particle physics use case is $10^6$. Horizontal axes indicate a measure defining the network complexity/size; this is the hidden layer density $\alpha$ for the condensed matter physics use case and the output dimension $z$ for the particle physics use case.}
    \label{fig:panel}
\end{figure*}

\section{\label{sec:results}Results}
\subsection{Condensed Matter Physics}
Figure \ref{fig:panel}a shows the average energy of inference and \ref{fig:panel}b the throughput, both as function of the hidden layer density $\alpha$. For the NVIDIA GPU the energy consumption (throughput) increases (decreases) with hidden layer size and by increasing the data precision from floating point 32 to floating point 64, two commonly used precision settings in the field. For both metrics the GPU outperforms the CPU up to a factor 10. The dual socket Intel Xeon Gold CPU features a non-monotonic dependence on network size and floating point precision. This is attributed to irregular utilization of the cores (40 in total). Forcing execution on a single core recovers the expected linear scaling with hidden layer density as we have confirmed by independent measurements (data not shown).

The energy and throughput estimates from the proposed MP-AIMC architecture are also shown in Figure \ref{fig:panel}a and \ref{fig:panel}b (dashed lines). The estimated energy use of this architecture is up to a factor $10^{3}$ lower than the CPU and GPU.
We observe that the MP-AIMC design yields a throughput comparable to CPU. However, GPU outperforms the MP-AIMC throughput up to a factor 7 for $\alpha\leq4$. For large networks the GPU throughput reduces whereas MP-AIMC remains flat by design, given that the hidden layer fits into a single MP-AIMC tile. This suggests that for even larger networks MP-AIMC will have an advantage also for the throughput. In addition, we note that unlike both the CPU and the GPU, the proposed MP-AIMC architecture features no parallelisation. The throughput of the MP-AIMC architecture naturally increases if multiple networks are run in parallel using more tiles and DPU resources at the expense of an increase in area of the chip.

\subsection{Particle Physics}
Figure \ref{fig:panel}c shows the average energy consumption of a batch of inferred events and Figure \ref{fig:panel}d the throughput as function of the individual network design of a single SVDD in the SVDD ensemble, labeled with $z$, the dimension of the manifold. Distinct from the condensed matter physics use case, GPU and CPU score very similar results for both metrics, in particular for single precision. Interestingly, in this case, the CPU outperforms the GPU for the energy cost at the largest network sizes ($z>144$). Also for the throughput, the performance difference between CPU and GPU is small. GPU is found to be superior to the CPU for double precision, while for single precision again CPU performs better, now for all network sizes. This nontrivial behavior is attributed to the computational cost stemming from the distance calculation, Eq.~\eqref{eqn:distancecalculation}, which in turn determines the anomaly score, which we confirmed by leaving out the distance calculation in the code. 

Figure \ref{fig:panel}c and \ref{fig:panel}d also includes the energy estimation of the proposed MP-AIMC architecture, which is observed to be up to a factor of $10^{3}$ more efficient than CPU and GPU. By the pipelined dataflow, the MP-AIMC throughput is the same as for condensed matter physics use case. However, on GPU and CPU the inference with the more complex and deeper structure of the SVDD requires more computational efforts than evaluation of the shallow RBM. As a result, MP-AIMC can yield over a factor $20$ faster throughput than CPU and GPU at the largest SVDD output dimensions.

\section{Conclusion}
We have presented a methodology to measure energy cost and compute time on CPU and GPU for inference tasks based on ANNs. By applying this methodology to NQS for quantum many-body systems in condensed matter physics and SVDD networks for anomaly detection in particle physics, we found that benefits of GPU as compared to CPU for energy efficiency and throughput strongly depend on the ANN architecture and on non-MVM operations. In particular, CPU can outperform GPU even for the largest networks considered. Therefore, energy benchmarks are always important, especially when working with unorthodox experimental ANN-based algorithms. 

Furthermore, we have proposed a dedicated MP-AIMC architecture capable of implementing both physics use cases, based on which the energy consumption and throughput can be estimated. By comparing the measured energy on CPU and GPU with the energy estimations for MP-AIMC, it is found that the latter improves the energy efficiency up to a factor $10^3$ for both the condensed matter physics and particle physics use case. The throughput is flat as a function of the network size in the proposed MP-AIMC architecture, as long as the network fits in the $512\times512$ crossbar array. For the relatively small networks used for NQS tested for the condensed matter physics use case, this yields a MP-AIMC throughput comparable with that of the CPU, whereas GPU throughput is up to factor 7 higher. Importantly, for the larger SVDD networks in the particle physics use case, the MP-AIMC throughput is always higher, over a factor 20 for the largest network tested. 

The benchmarks performed suggest great potential for neuromorphic accelerators based on MP-AIMC. A key challenge associated with AIMC is computational imprecision\cite{le2018mixed}. 
Yet, solutions on device\cite{giannopoulos20188}, unit cell and circuit\cite{le2022precision}, as well as algorithmic level\cite{joshi2020accurate} have been shown to be effective in compensating for the low-precision analog computing. It is foreseen that similar approaches can be effective for the networks presented in this work and consider studying the impact of precision as a very interesting topic for future work. 
Moreover, future work may focus on next-generation cross-bar structures and parallel architectures which can further improve the throughput. Combined with the fundamentally flat scaling of the compute time with the network size, this suggest great potential for scientific workloads with exceptionally high inference demands, potentially enabling so far uncomputable  tasks in scientific computing.

\section*{Acknowledgements}
We thank Giammarco Fabiani for stimulating discussions and support on the condensed matter physics use case, as well as for providing the trained networks, Julita Corbalan for support on operating the EAR software package,
We thank our colleagues at IBM Research, in particular Y. Kohda, S. Munetoh and G. Karunaratne for discussions on the MP-AIMC power/performance analysis and acknowledge the support from the IBM Research AI Hardware Center. 
This research received funding from the Interdisciplinary Research Platform (IRP) for Green Information Technology at the Faculty of Science of Radboud University that aims to promote pioneering research ideas in emerging interdisciplinary themes, the Nederlandse Organisatie voor Wetenschappelijk Onderzoek (NWO) and is part of the Shell NWO/FOM inititative “Computational sciences for energy research” of Shell and Chemical Sciences, Earth and Life Sciences, Physical Sciences, Stichting voor Fundamenteel Onderzoek der Materie (FOM) and Stichting voor de Technische Wetenschappen (STW). It also received funding from the European Research Council under ERC Grant Agreement No. 856538 (3D-MAGiC), European Union's Horizon Europe research and innovation programme under Grant Agreement No. 101046878 (HYBRAIN), the State Secretariat for Education, Research and Innovation (SERI) No. 22.00029 and we gratefully acknowledge the computer resources at Artemisa, funded by the European Union ERDF and Comunitat Valenciana as well as the technical support provided by the Instituto de Fisica Corpuscular, IFIC (CSIC-UV). R. RdA also acknowledges the Ministerio de Ciencia e Innovación (PID2020-113644GB-I00).

\nocite{*}
\bibliography{aipsamp}

\end{document}